\documentclass[aps,pra,twocolumn,superscriptaddress,amsmath,amssymb]{revtex4}
\usepackage{graphicx}
\usepackage{amsmath}
\usepackage{color}
\usepackage[normalem]{ulem}
\def\bm#1{\mathbf{#1}}

\begin{document}

\title{Repulsive polarons in alkaline-earth(-like) atoms across an orbital Feshbach resonance}

\author{Tian-Shu Deng}
\affiliation{Key Laboratory of Quantum Information, University of Science and Technology of China, Chinese Academy of Sciences, Hefei, Anhui, 230026, China}
\affiliation{Synergetic Innovation Center of Quantum Information and Quantum Physics, University of Science and Technology of China, Hefei, Anhui 230026, China}
\author{Zhuo-Cheng Lu}
\affiliation{Department of Physics, Renmin University of China, Beijing 100872, China}
\author{Yue-Ran Shi}
\affiliation{Department of Physics, Renmin University of China, Beijing 100872, China}
\author{Jin-Ge Chen}
\affiliation{Department of Physics, Renmin University of China, Beijing 100872, China}
\author{Wei Zhang}
\email{wzhangl@ruc.edu.cn}
\affiliation{Department of Physics, Renmin University of China, Beijing 100872, China}
\affiliation{Beijing Key Laboratory of Opto-electronic Functional Materials and Micro-nano Devices,
Renmin University of China, Beijing 100872, China}
\author{Wei Yi}
\email{wyiz@ustc.edu.cn}
\affiliation{Key Laboratory of Quantum Information, University of Science and Technology of China, Chinese Academy of Sciences, Hefei, Anhui, 230026, China}
\affiliation{Synergetic Innovation Center of Quantum Information and Quantum Physics, University of Science and Technology of China, Hefei, Anhui 230026, China}

\date{\today}
\begin{abstract}
We characterize properties of the so-called repulsive polaron across the recently discovered orbital Feshbach resonance in alkaline-earth(-like) atoms. Being a metastable quasiparticle excitation at the positive energy, the repulsive polaron is induced by the interaction between an impurity atom and a Fermi sea. By analyzing in detail the energy, the polaron residue, the effective mass, and the decay rate of the repulsive polaron, we reveal interesting features that are intimately related to the two-channel nature of the orbital Feshbach resonance. In particular, we find that the life time of the repulsive polaron is non-monotonic in the Zeeman-field detuning bewteen the two channels, and has a maximum on the BEC-side of the resonance. Further, by considering the stability of a mixture of the impurity and the majority atoms against phase separation, we show that the itinerant ferromagnetism may exist near the orbital Feshbach resonance at appropriate densities. Our results can be readily probed experimentally, and have interesting implications for the observation of itinerant ferromagnetism near an orbital Feshbach resonance.
\end{abstract}
\maketitle

\section{Introduction}

The recently discovered orbital Feshbach resonance (OFR) in $^{173}$Yb opens up the avenue of investigating strongly-interacting many-body physics using alkaline-earth(-like) atoms~\cite{ren1,ofrexp1,ofrexp2}. In an OFR, the spin-exchange interaction between the ground $^1S_0$ and the long-lived excited $^3P_0$ hyperfine manifolds can be tuned by an external magnetic field. It follows that the wealth of precision quantum control techniques, which have been developed for the purpose of quantum metrology and quantum information using the clock-state manifolds ($\{^1S_0,^3P_0\}$), can be employed to engineer highly non-trivial many-body scenarios~\cite{AE1,AE2,AEnew7,AE3,AEnew3,AE4,congjun03,AE5,AEnew1,AEnew4,AEnew5,phasePRL,phaseEPL,quella13,ofr1,ofr2,ofr3,AEnew8,phases1,quella15,AEnew9,phases2,phases3,ye2016old,ye2016,fallani2016,gyuboong,ofrwidth}. Recent studies in this regard range from interaction-induced topological states~\cite{zhou,Zoller}, to impurity problems such as the Kondo effects~\cite{kondo1,kondo2,kondo3,kondo4,kondo5,kondo6} and the polaron to molecule transitions~\cite{OFRPol,qiranpol}. Naturally, the key properties of these phenomena are firmly based on the features of interactions of an OFR.

Like the interactions of Feshbach resonance in alkali atoms, the interactions of OFR can be understood as the resonant scattering between an open and a closed channel. Consider two alkaline-earth(-like) atoms respectively in the $^1S_0$ (denoted as $|g\rangle$) and the $^3P_0$ (denoted as $|e\rangle$) manifolds, as $J=0$ for these so-called clock-state manifolds, the nuclear and the electronic spin degrees of freedom are decoupled. Denoting a particular nuclear spin state $m_I$ ($m_{I+1}$) in each manifold as $|\uparrow\rangle$ ($|\downarrow\rangle$), we may associate the open channel with the $|g\downarrow\rangle$ and $|e\uparrow\rangle$ states, and the closed channel with the $|g\uparrow\rangle$ and $|e\downarrow\rangle$ states. Due to the differential Zeeman shift in the clock-state manifolds~\cite{zeemanshift1,zeemanshift2}, an external magnetic field can conveniently shift the detuning between the open- and the closed-channel scattering thresholds. Further, as the short-range interaction of the OFR occurs either in the electronic spin-singlet and nuclear spin-triplet channel, or the electronic spin-triplet and nuclear spin-singlet channel, it couples the closed- and the open channels together. The scattering resonance occurs when the energy of a bound state in the closed channel is tuned to the open-channel scattering threshold. As a result of OFR, a crossover from the Bardeen-Cooper-Schrieffer (BCS) to the Bose-Einstein condensation (BEC) regime can be realized in alkaline-earth(-like) atoms by tuning the magnetic field, which is similar to the magnetic Feshbach resonance in alkali atom. However, the existence of multiple nuclear spin states, as well as the spin-exchange interactions in the OFR complicate the two-body scattering process, and lead to rich physics in the many-body setting.

An illuminating example here is the system consisting of a mobile impurity interacting with its environment. As the limiting case of a many-body system in the large polarization limit, mobile impurity and its associated quasi-particle excitations contain valuable information of the underlying system. Whereas impurity problems in the background of Bose gases or Fermi condensates have attracted considerable attention in recent years~\cite{bp1,bp2,bp3,bp4,bp45,bp5,bp6,bp7,bp8,bp9,bp10,bp11,bp12,bp13,bp14,bp15,bp16,bp17,fermicondpolaron1,fermicondpolaron2}, here we focus on the case of an impurity against a non-interacting Fermi sea. In alkali atoms, it has been shown that the impurity can either form a tightly bound molecule with a majority atom, or induce collective particle-hole excitations in the Fermi sea and form the so-called Fermi polaron~\cite{mitpolaron,enspolaron,grimmpolaron,kohlpolaron,impurityreview1,impurityreview2}. A polaron to molecule transition has been observed experimentally, as the interaction is tuned. Further, at positive energies, a so-called repulsive polaron branch exists, which is metastable and associated with the elusive itinerant ferromagnetism~\cite{grimmpolaron,kohlpolaron,reppol1,reppol2,reppol3,reppol4,reppol7,reppol8}. In OFR, a recent theoretical study suggests that the transition between the attractive polaron and the molecule also exists when tuning the magnetic field~\cite{OFRPol,qiranpol}. However, the existence and properties of the repulsive polaron branch have not been investigated.

In this work, we characterize properties of the repulsive polaron across the OFR, using the parameters of $^{173}$Yb atoms as a concrete example. As illustrated in Fig.~\ref{fig:fig1s}, we consider a single impurity atom in the $|e\uparrow\rangle$ state, which interacts with a Fermi sea of atoms in the $|g\downarrow\rangle$ state. While the impurity and the background atoms are initially in the open channel, the spin-exchange interactions would scatter atoms into the closed channel. Adopting the T-matrix formalism~\cite{reppol3,reppol4}, we demonstrate the existence of a metastable repulsive polaron branch at positive energies across the OFR. We characterize various properties of the repulsive polaron, such as the energy, the polaron residue, the effective mass, and the decay rate. In particular, we identify unique features in all of these quantities, which are intimately related to the two-channel nature of the OFR. An interesting result of the inter-channel scattering is that the life time of the repulsive polaron is non-monotonic in the effective interaction strength, and has a maximum on the BEC-side of the resonance. We further analyze the condition for the existence of itinerant ferromagnetism in these atoms near an OFR. By considering the stability of a homogeneous mixture of the impurity $|e,\uparrow\rangle$ atoms and the majority $|g,\downarrow\rangle$ atoms against phase separation, we show that a phase-separated state, and hence the itinerant ferromagnetism, can be stabilized beyond a critical Zeeman-field detuning. Since such a conclusion is conditional on the stability of the repulsive polaron, we further demonstrate that for appropriate atomic densities, a parameter window exists where the system favors phase separation and the repulsive polaron is long-lived and away from the molecule-hole continuum. Our findings can be readily probed experimentally, and have interesting implications for the observation of itinerant ferromagnetism near an OFR.

The paper is organized as follows. In Sec.~\ref{sec_form}, we present the T-matrix formalism for Fermi polarons in the context of an OFR. We demonstrate the existence of the repulsive polaron, and characterize its energy by calculating the spectral function in Sec.~\ref{sec_spectral}. We then study in detail the polaron residue and the effective mass in Sec.~\ref{sec_prop}, where kinks in these properties are identified and associated with resonant scatterings in the many body background. We characterize the decay rate of the repulsive polaron in Sec.~\ref{sec_decay}, and discuss in detail the potential stability region of the itinerant ferromagnetism near an OFR in Sec.~\ref{sec_ps}. Finally, we summarize in Sec.~\ref{sec_sum}.

\section{T-matrix formalism}\label{sec_form}
\begin{figure}[tbp]
\centering
\includegraphics[width=8cm]{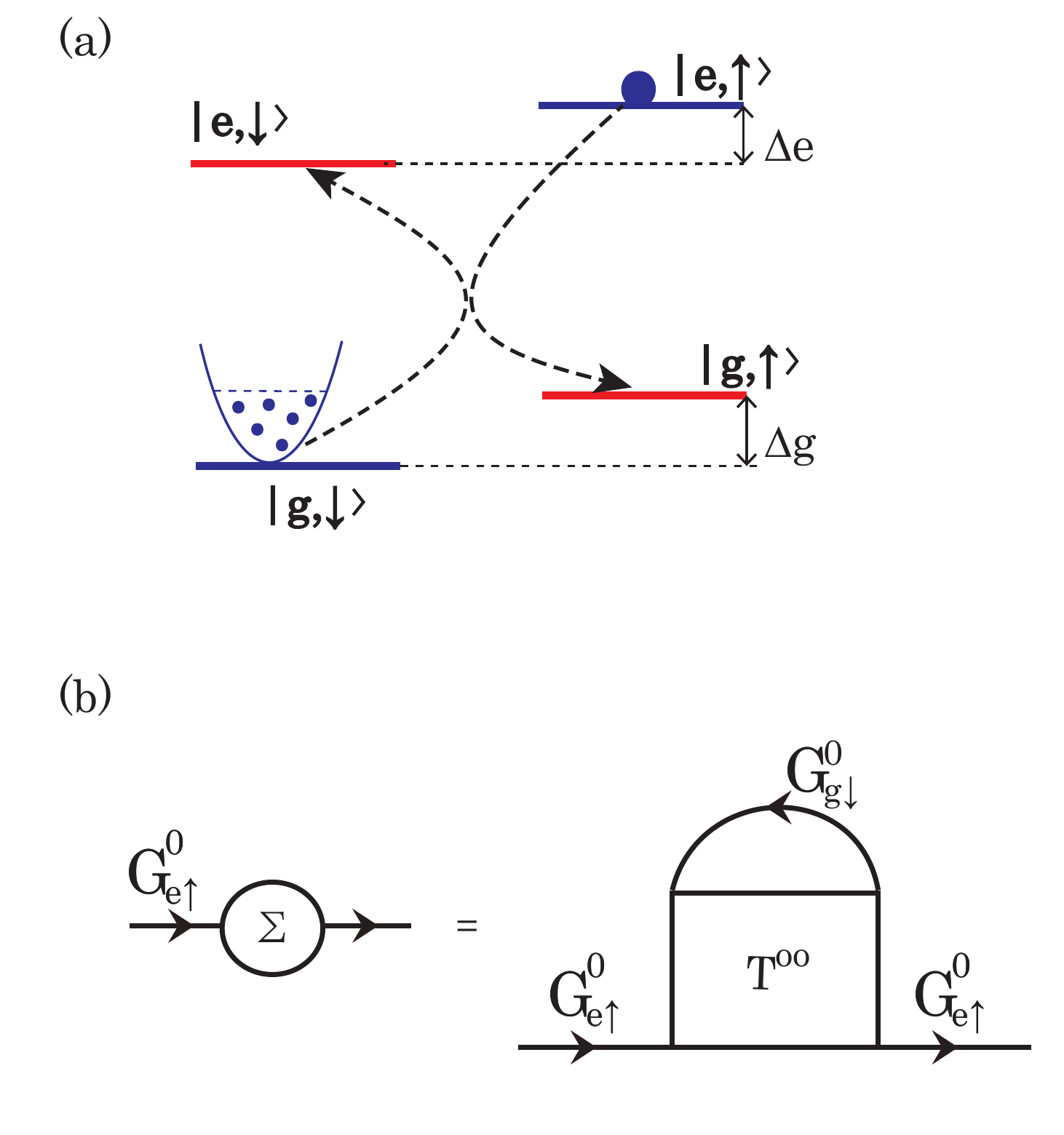}
\caption{(a) Level diagram of an OFR in alkaline-earth(-like) atoms. An impurity of $|e,\uparrow\rangle$ is immersed in a majority Fermi sea of $|g,\downarrow\rangle$ atoms, and can be scattered to the other two atomic states forming the closed channel via interaction. $\Delta_g={g_g} \mu_B B$ and $\Delta_e={g_e} \mu_B B$ are the Zeeman shifts of the $|g\rangle$ and $|e\rangle$ manifolds, respectively. (b) The one-hole polaron self-energy $\Sigma$ near an OFR. The solid lines with arrows indicate free propagator $G^0$ for $|g,\downarrow\rangle$ or $|e,\uparrow\rangle$, and the square $T^{oo}$ indicates the T-matrix with the incoming and the outgoing states being both in the open channel.}
\label{fig:fig1s}
\end{figure}

We start from the non-interacting Hamiltonian corresponding to the configuration in Fig.~\ref{fig:fig1s}(a)
\begin{align}\label{H0}
H_{0} & =\sum_{\bm{k}}\epsilon_{\bm{k}}^{o}(a_{g,\downarrow\bm{k}}^{\dagger}a_{g,\downarrow\bm{k}}+a_{e,\uparrow,\bm{k}}^{\dagger}a_{e,\uparrow,\bm{k}})\nonumber\\
&+\sum_{\bm{k}}\epsilon_{\bm{k}}^{c}(a_{e,\downarrow\bm{k}}^{\dagger}a_{e,\downarrow\bm{k}}+a_{g,\uparrow\bm{k}}^{\dagger}a_{g,\uparrow\bm{k}}),
\end{align}
where $a_{j,\sigma,\bm{k}}^\dagger$ ($a_{j,\sigma,\bm{k}}$) creates (annihilates) an atom in the corresponding pseudo-spin state $|j,\sigma\rangle$ ($j \in \{g,e\}$, $\sigma \in \{\uparrow,\downarrow\}$) with momentum $\bm{k}$. Here, $\epsilon_{\bm{k}}^{o}=\hbar^{2}\bm{k}^{2}/2m$ and $\epsilon_{\bm{k}}^{c}=\hbar^{2}\bm{k}^{2}/2m+\delta/2$. The detuning between the two channels $\delta \equiv \Delta_g - \Delta_e =(g_g-g_e)\mu_B B$ originates from the differential Zeeman shift of the clock states in the presence of a magnetic field $B$, where $g_g$ ($g_e$) is the Lande $g$-factor for the $|g\rangle$ ($|e\rangle$) manifold, and $\mu_B$ is the Bohr magneton.

The typical inter-orbital spin-exchange interaction of an OFR can be written as
\begin{equation}\label{Hint}
    H_{\rm int}=\frac{g_{+}}{2}\sum_{\bm{q}}A_{+}^{\dagger}(\bm{q})A_{+}(\bm{q})+\frac{g_{-}}{2}\sum_{\bm{q}}A_{-}^{\dagger}(\bm{q})A_{-}(\bm{q}),
\end{equation}
where we have
\begin{align}\label{Apm}
A_{\pm}(\bm{q})=\sum_{\bm{k}}(a_{e,\downarrow,\bm{k}}a_{g,\uparrow,\bm{q-k}}{\mp}a_{e,\uparrow,\bm{k}}a_{g,\downarrow,\bm{q-k}}),
\end{align}
and the interaction strengths $g_\pm$ are related to the physical ones via the renormalization relation $1/g_{\pm}=1/\tilde{g}_{\pm}-\sum_{\bm{k}}1/2\epsilon_{\bm {k}}^o$ with $\tilde{g}_{\pm}={4\pi\hbar^{2}a_{\pm}}/{m}$. Throughout this work, we adopt the parameters of $^{173}$Yb atoms, with $a_+=1900a_0$ and $a_-=219.5a_0$~\cite{ofrexp1,ofrexp2,huhui,defecttheory}.

Diagrammatically, the polaron properties can be calculated using the retarded self-energy of the impurity atom~\cite{reppol3}, which is given by [see Fig.~\ref{fig:fig1s}(b)]
\begin{align}\label{Sig}
  \Sigma(\bm{Q},E)=\int\frac{d\bm{q}}{(2\pi)^{3}}\int\frac{d\omega}{2\pi}G_{g\downarrow}^{0}(\bm{q},\omega)T^{oo}(\bm{q+Q},E+\omega),
\end{align}
where $G_{g,\downarrow}^{0}(\bm{q},\omega)=({\omega+i0^{+}-\epsilon_{\bm{q}}^{o}})^{-1}$ is the free-fermion propagator of the majority atoms, and $T^{oo}$ is the T-matrix describing the open-channel scattering processes. Here $E$ and $\bm{Q}$ are respectively the energy and the center-of-mass momentum of the self-energy, and $\omega$ is the Matsubara frequency. Due to the spin-exchange nature of the interaction, the open- and the closed-channel scattering matrices are coupled. Accordingly, there should be four kinds of T-matrices $T^{oo}$, $T^{oc}$, $T^{co}$, and $T^{cc}$, with the incoming and the outgoing states being in either the open or the closed channel, as indicated by the superscript labels. As discussed in Ref.~\cite{ofrwidth}, under the ladder approximation, we may write down a set of coupled equations for the T-matrices, which lead to the solution
 \begin{equation}\label{Too}
T^{oo}(\bm{q},\omega)=\frac{\frac{1}{2}(g_{+}+g_{-})-g_{+}g_{-}\chi^{c}}{1-\frac{1}{2}(g_{+}+g_{-})(\chi^{o}+\chi^{c})+g_{+}g_{-}\chi^{o}\chi^{c}}.
 \end{equation}
Here, the pair propagators for the closed and the open channel $\chi^{c}(\bm{q},\omega)$ and $\chi^{o}(\bm{q},\omega)$ can be written as
  \begin{align}
    \chi^{c}(\bm{q},\omega)&=\sum_{\bm{k}}\frac{1}{\omega+i0^{+}-\epsilon_{\bm{k}}^{c}-\epsilon_{\bm{q}-\bm{k}}^{c}},\label{chic}\\
  \chi^{o}(\bm{q},\omega)&=\sum_{|\bm{k}|>k_{F}}\frac{1}{\omega+i0^{+}-\epsilon_{\bm{k}}^{o}-\epsilon_{\bm{q}-\bm{k}}^{o}},\label{chio}
\end{align}
where the Fermi wave vector $k_F$ is related to the Fermi energy $E_F$ of $|g,\downarrow\rangle$ atoms as $E_F=\hbar^2k_F^2/2m$. From the equations above, we see that $\chi^{o}(\bm{q},\omega)$ and $\chi^{c}(\bm{q},\omega)$ are isotropic in $\bm{q}$. For the convenience of discussion, we define $q=|\bm{q}|$.

Substituting Eq. (\ref{Too}) into (\ref{Sig}), we obtain

\begin{align}\label{Sig2}
  &\Sigma(\bm{Q},E)  =\sum_{q<k_{F}} \bigg[ \frac{1}{2}\left(\frac{1}{\tilde{g}_{+}}+\frac{1}{\tilde{g}_{-}}\right)-\tilde{\chi}^{o}({\bm{q+Q}},E+\epsilon_{\bm{q}}^{o}) \nonumber\\
  &-\frac{1}{4}\left(\frac{1}{\tilde{g}_{+}}-\frac{1}{\tilde{g}_{-}}\right)^{2}\frac{1}{\frac{1}{2}(\frac{1}{\tilde{g}_{+}}+\frac{1}{\tilde{g}_{-}})-\tilde{\chi}^{c}({\bm{q+Q}},E+\epsilon_{\bm{q}}^{o})} \bigg]^{-1},
\end{align}
where the renormalized pair propagator $\tilde{\chi}^{c}=\chi^c+\sum_{\bm{k}}1/2\epsilon_{\bm {k}}^o$ and $\tilde{\chi}^{o}=\chi^o+\sum_{\bm{k}}1/2\epsilon_{\bm {k}}^o$. With Dyson's equation, the Green's function of an impurity $|e,\uparrow\rangle$ dressed by a Fermi sea of $|g\downarrow\rangle$ atoms can therefore be written as
\begin{equation}\label{GF}
  G_{e\uparrow}(\bm{Q},E)=\frac{1}{E+i0^{+}-\epsilon_{\bm{Q}}^{o}-\Sigma(\bm{Q},E)},
\end{equation}
from which we may extract various properties of the quasi-particle excitations.

\section{Spectral function and the polaron energy}\label{sec_spectral}
\begin{figure}[tbp]
\centering
\includegraphics[width=8cm]{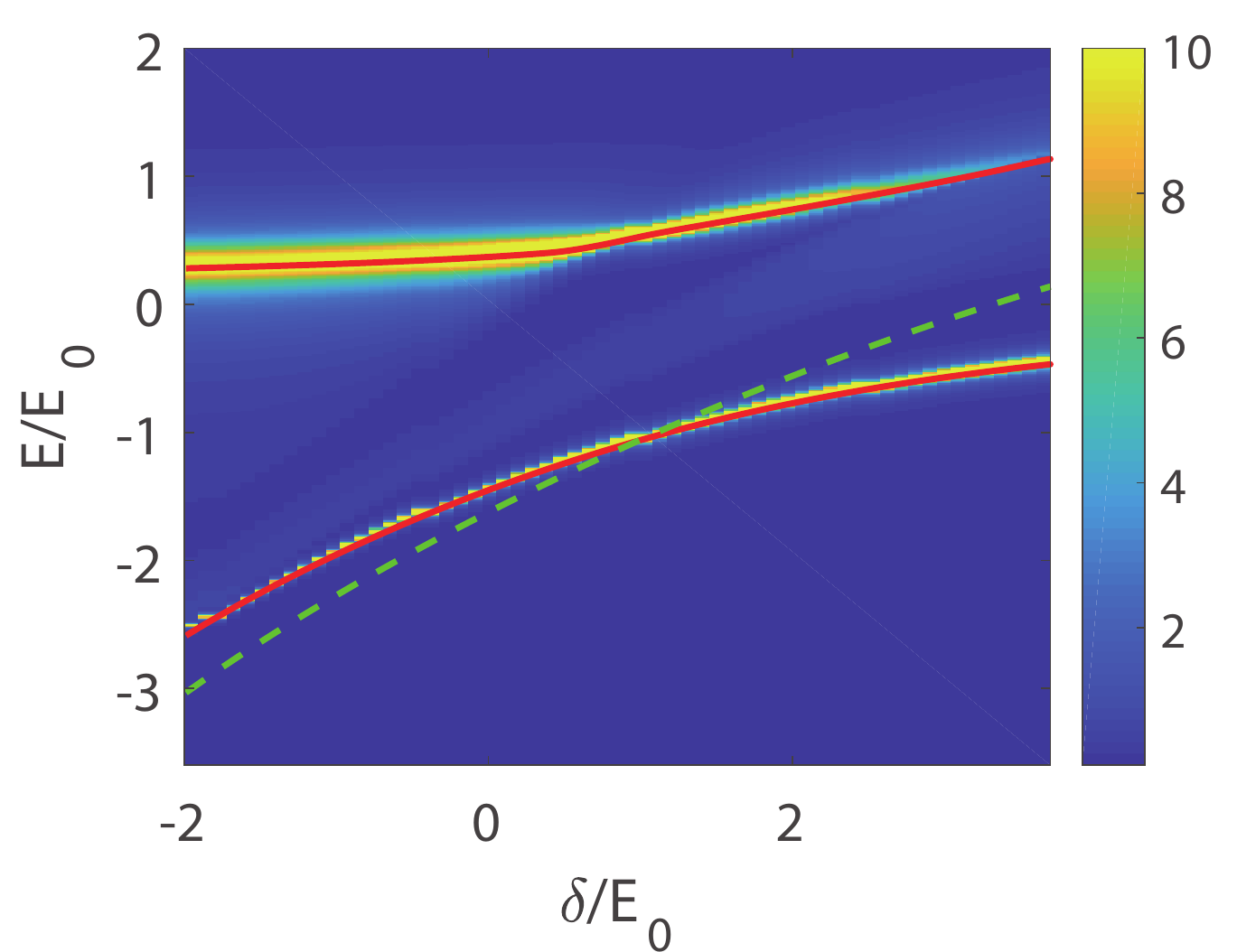}
\caption{False color plot of the spectral function $A(\bm{Q}=0,E)$ of an impurity $|e,\uparrow\rangle$ in a Fermi sea of non-interacting $|g\downarrow\rangle$ particles on the $\delta$--$E$ plane. The solid red lines depict the polaron energies given by Eq. (\ref{SigE}), and the dashed green line is the molecular energy. The light-blue area between the two polaronic branches is the molecule-hole continuum. The upper repulsive polaron branch merges into the molecule-hole continuum for $\delta\gtrsim 3.8E_0$. Here, we define the unit of energy $E_0=\hbar^2 k_0^2/2m$, where the unit Fermi wave vector $k_0^3=6\pi^2n_0$ and the unit density $n_0=5\times10^{13} {\rm cm}^{-3}$. In this plot, we take $n=n_0$.}
\label{fig:fig2s}
\end{figure}

\begin{figure}[tbp]
\centering
\includegraphics[width=8cm]{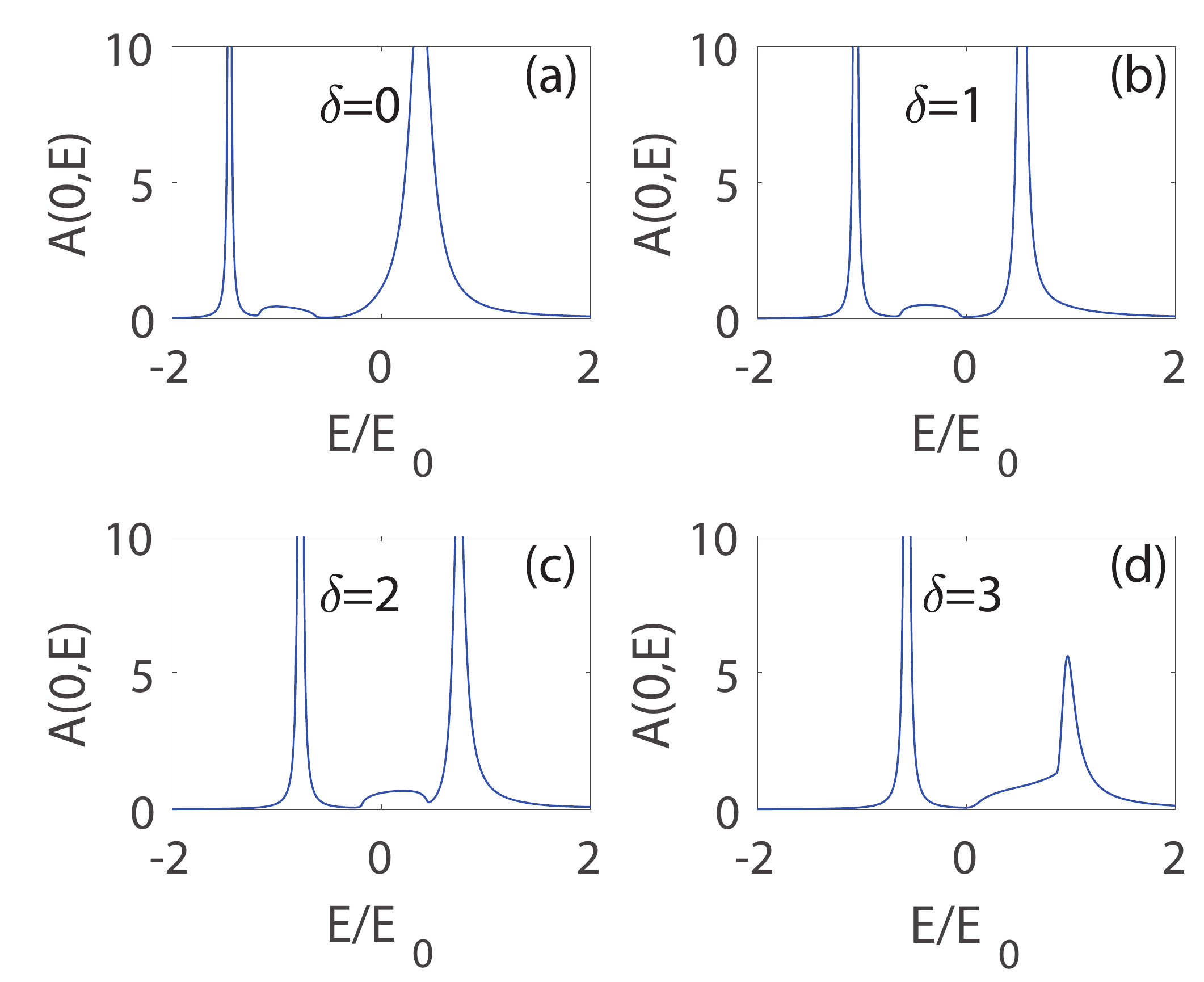}
\caption{The spectral function $A(\bm{Q}=0,E)$ as functions of $E$ with different detunings. We have taken the same parameters as those in Fig.~\ref{fig:fig2s}.}
\label{fig:fig3s}
\end{figure}

We first calculate the spectral function at zero temperature ($T=0$)
\begin{equation}\label{Aw}
  A(\bm{Q},E)=-2{\rm Im}G_{e\uparrow}(\bm{Q},E).
\end{equation}
In Fig.~\ref{fig:fig2s}, we plot $A(\bm{Q}=0, E)$ in the $\delta$--$E$ plane. The spectral function is strongly peaked at the energies of polaron excitations satisfying
\begin{equation}\label{SigE}
  E_{\pm}={\rm Re[}\Sigma(\bm{Q},E_{\pm})].
\end{equation}
As is apparent in Fig.~\ref{fig:fig2s}, there exist two solutions for Eq. (\ref{SigE}). The lower branch with $E = E_-<0$ corresponds to the attractive polaron, and the upper branch with $E = E_+>0$ corresponds to the repulsive polaron. In contrast to the attractive polaron, which is undamped under the ladder approximation here, the repulsive polaron, being a mestastable quasipaticle excitation with $E_+>0$, features a finite width in the spectral function as illustrated in Fig.~\ref{fig:fig3s}, which originates from the decay into low-lying states. Under the inter-orbital spin-exchange interactions of the OFR, as we will show later, the finite spectral width and hence the decay of the repulsive polaron mainly come from the resonant coupling of the quasi-particle excitation to the open- and the closed-channel scattering continuum. Finally, we notice the existence of a broad wing between the two polaron peaks as shown in Fig~\ref{fig:fig3s}, which corresponds to the molecule-hole continuum.

\section{Impurity residue and the effective mass}\label{sec_prop}
We now characterize the impurity residue and the effective mass of the repulsive polaron. For a polaron excitation, the quasi-particle residue is defined as~\cite{reppol3}
\begin{equation}\label{Z}
  Z_{\pm}=\frac{1}{1-{\rm Re}\left[\frac{\partial\Sigma(0,\omega)}{\partial\omega}\right]}\Bigg|_{\omega=E_{\pm}},
\end{equation}
and its effective mass as
\begin{equation}\label{mstar}
  m_{\pm}^{*}=\frac{1}{Z_{\pm}}\frac{1}{1+{\rm Re}\left[\frac{\partial\Sigma(\bm{Q},\omega)}{\partial \bm{Q}^{2}}\right]}
  \Bigg|_{\bm{Q}=0,\omega=E_{\pm}},
\end{equation}
where the subscript $+$ ($-$) labels the repulsive (attractive) branch of polarons.

\begin{figure}[tbp]
\centering
\includegraphics[width=9cm]{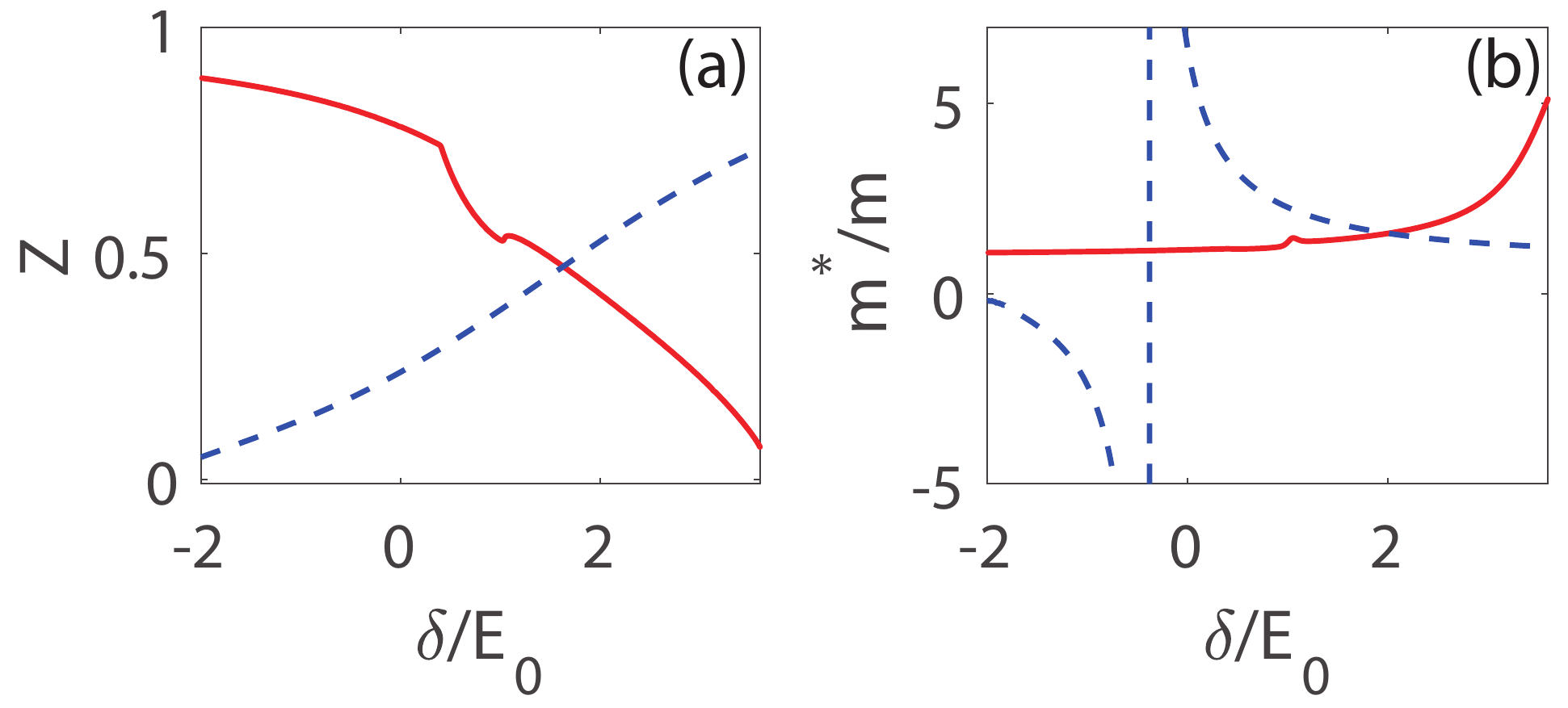}
\caption{(a) Quasi-particle residues $Z_{\pm}$ for the attractive (blue dashed) and repulsive (red solid) polarons as functions of $\delta$. (b) Effective masses of the attractive (blue dashed) and the repulsive (red solid) polarons as functions of $\delta$.}
\label{fig:fig4s}
\end{figure}

We have shown the quasi-particle residue as well as the effective mass of the repulsive polaron in Fig.~\ref{fig:fig4s}. For comparison, we have also plotted the residue and the effective mass of the attractive polaron. In an OFR and under the setup illustrated in Fig.~\ref{fig:fig1s}, the resonance occurs at $\delta_0 \sim 3.06 E_0$, and the system is on the BCS-side of the resonance for $\delta>\delta_0$. In Fig.~\ref{fig:fig4s}, we see that as $\delta$ increases (i.e., moves towards the BCS side of the resonance), $Z_+$ decreases and $m_+^*$ increases, which are qualitatively consistent with the case of alkali atoms. A prominent difference in the current case is the existence of kinks in both the residue and the effective mass at $\delta = E_+$ and $\delta = E_+ + E_F/2$. The occurrence of these kinks can be explained by the qualitative difference, between regions with different values of $\delta$, in the way that the atoms in the open-channel Fermi sea are scattered into the closed-channel
continuum in forming the repulsive polaron.

The location of the kinks can be determined analytically by considering the scattering process between the impurity and the majority atoms, in which the out-going states are at the closed-channel scattering threshold. In particular, because $\delta$ represents the closed-channel detuning of the two atoms and $E_+$ is the interaction-induced energy shift of the impurity atom, at $\delta=E_+$ an impurity atom with zero momentum can interact with a majority atom at the bottom of the Fermi sea (with $q=0$), which are resonantly scattered to two atoms in the closed-channel scattering threshold. Likewise, at $\delta=E_+ +E_F/2$, an impurity atom with zero momentum interacts with a majority atom on the Fermi surface (with $q=k_F$), which, under the momentum conservation, are resonantly scattered to two atoms in the closed channel each with a momentum $q=k_F/2$. The process can therefore be qualitatively described as resonant scatterings in the many-body background.

To further demonstrate this point, in Fig.~\ref{fig:fig5s}(a) and \ref{fig:fig5s}(b), we explicitly show the imaginary parts of the pair propagators in the open- and the closed-channel, respectively, on the $\delta$--$q$ plane. As the imaginary parts of the pair propagators are related to the removable singularities in the summation of Eqs.~(\ref{chic}) and (\ref{chio}), they reflect the contribution to the polaron self-energy as atoms in the Fermi sea with momentum $q<k_F$ is resonantly coupled to the scattering states in the open(closed)-channel continuum, forming particle-hole excitations. In the case of the open-channel pair propagator, for any given $\delta$, a finite imaginary part exists only when the magnitude of the center-of-mass momentum of the hole excitation $q$ is below a critical value. This implies that, for any given $\delta$, part of the Fermi sea is blocked by the energy and momentum conservation conditions such that atoms therein cannot be resonantly scattered into the open-channel continuum. The case of the closed-channel pair propagator is more complicated. For $\delta<E_+$, $\chi_c({\bm{q}},E_+ +\epsilon_{\bm{q}})$ features a finite imaginary part for all $q<k_F$. Therefore, all atoms in the Fermi sea can be resonantly scattered into the closed-channel continuum to form the particle-hole excitations. For $\delta>E_+ +E_F/2$, on the other hand, $\chi_c({\bm{q}},E_+ +\epsilon_{\bm{q}})$ can be completely real for any $q<k_F$. Hence, none of the atoms in the Fermi sea can be resonantly scattered into the closed-channel continuum, and the particle-hole excitations in the repulsive polaron is open-channel dominated. Therefore, in different regions of $\delta$, the closed-channel scattering continuum contribute in qualitatively different ways to the polaron self-energy, which gives rise to the appearance of kinks at the boundaries of these regions.

\section{Decay rate of the repulsive polaron}\label{sec_decay}
Being a mestable state, the repulsive polaron can decay into low-lying states.
Experimentally, it has been shown that for alkali atoms, the dominating decay channel for the repulsive polaron is the coupling to the bare impurity state in the attractive-polaron branch, so long as the interaction is not in the deep BEC regime~\cite{reppol8}. We assume that the case with alkaline-earth(-like) atoms is similar. One should then include the corresponding decay channel in the diagrams leading to the repulsive polaron self-energy. Such a decay rate can be calculated as~\cite{reppol8}
\begin{equation}\label{decay}
  \Gamma=-2Z_{+}[{\rm Im}\Sigma(0,E_{+})],
\end{equation}
where $Z_+$ is the residue for the repulsive polaron. Further, we can replace the free-fermion propagator $G_{e\uparrow}^{0}$ with $(1-Z_+)G_{e\uparrow}^{0}$ in the self-energy $\Sigma$, which implies substituting $\chi^o$ with $(1-Z_+)\chi^o$. This leads to $\Gamma=\sum_{|\bm{q}|<k_{F}}\Gamma({\bm{q}})$ with
\begin{align}\label{Sigq}
  \Gamma({\bm{q}})=&{\rm Im}\Big\{-2Z_{+}\Big[\frac{1}{2}\left({\tilde{g}_{+}^{-1}}+{\tilde{g}_{-}^{-1}}\right)
  \nonumber \\
  &-(1-Z_{+})\tilde{\chi}^{o}({\bm{q}},E_{+}+\epsilon_{\bm{q}}) \nonumber\\
  &-\frac{1}{4}
  \frac{\left({\tilde{g}_{+}^{-1}}-{\tilde{g}_{-}^{-1}}\right)^{2}}
  {\frac{1}{2}\left({\tilde{g}_{+}^{-1}}+{\tilde{g}_{-}^{-1}}\right)-\tilde{\chi}^{c}({\bm{q}},E_{+}+\epsilon_{\bm{q}})}\Big]^{-1}\Big\}.
\end{align}

\begin{figure}[tbp]
\centering
\includegraphics[width=9cm]{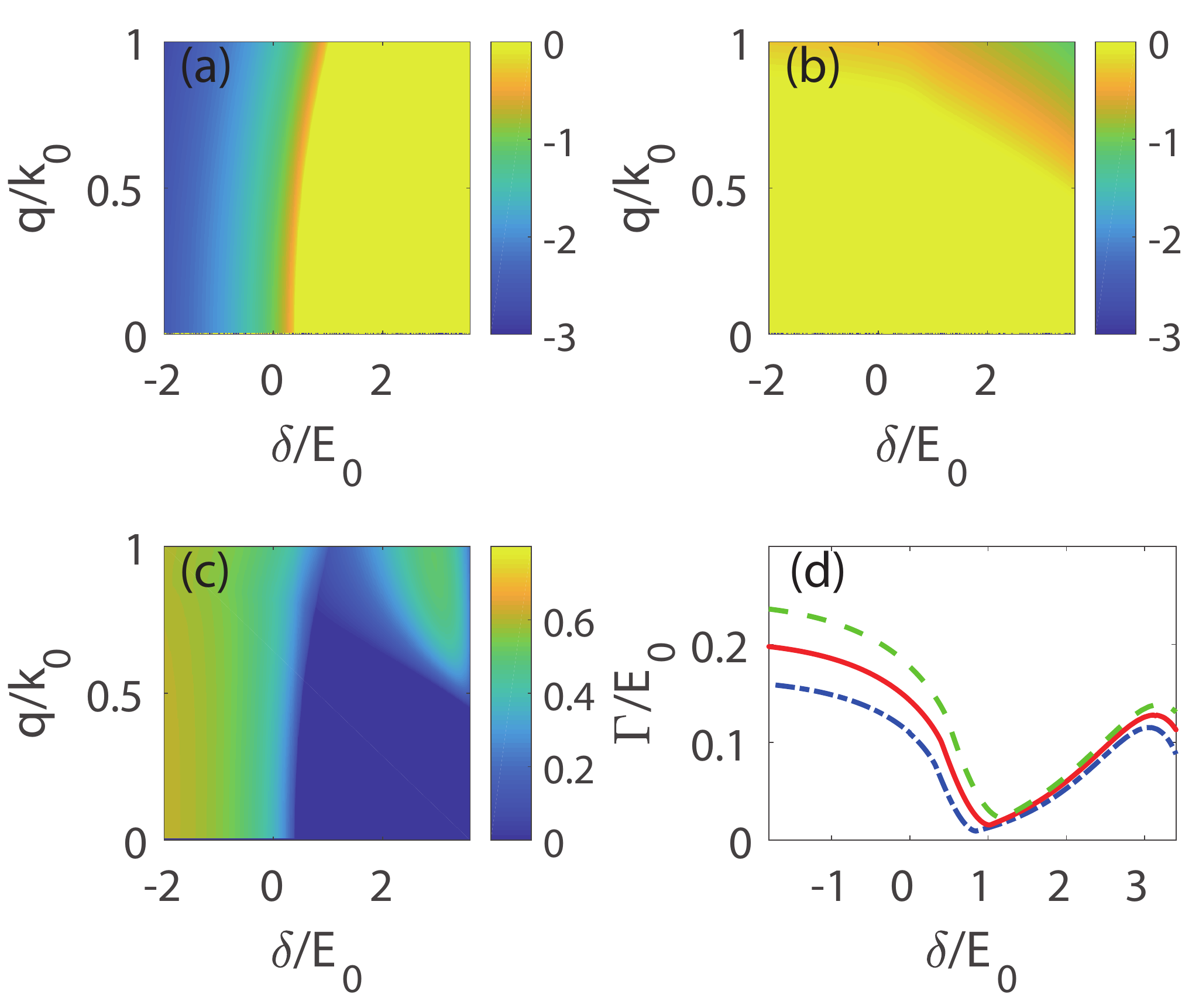}
\caption{(a) Imaginary part of the closed-channel pair propagator $\chi^{c}({\bm{q}},E_{+}+\epsilon_{\bm{q}}^{o})$ on the $\delta$--$q$ plane. (b)Imaginary part of the open-channel pair propagator $\chi^{o}({\bm{q}},E_{+}+\epsilon_{\bm{q}}^{o})$ on the $\delta$--$q$ plane. (c) $\Gamma({\bm{q}})$ on the $\delta$--$q$ plane. (d) The decay rate $\Gamma$ as function of $\delta$. In (a-c), we take $n= n_0$. In (d), the green dashed line, the red solid line, and the blue dash-dotted line correspond to cases of $n=1.2 n_0$, $n_0$, and $0.8 n_0$, respectively.}
\label{fig:fig5s}
\end{figure}

We plot the calculated $\Gamma({\bm{q}})$ and $\Gamma$ in Fig.~\ref{fig:fig5s}(c) and \ref{fig:fig5s}(d), respectively. In Fig.~\ref{fig:fig5s}(c), we see that the decay rate consists of two different contributions, which can be numerically associated with contributions from the pair propagators ${\tilde \chi}_o$ and ${\tilde \chi}_c$.
At large and positive $\delta$, the decay of the repulsive polaron is open-channel dominated, which increases as the system is tuned further towards the BCS side. At smaller or negative $\delta$, the closed-channel contribution becomes important, which increases as the system is tuned towards the BEC regime. While this result is consistent with the previous analysis of the pair propagators, the competition between the two channels gives rise to the non-monotonic behavior of $\Gamma$ as shown in Fig.~\ref{fig:fig5s}(d), where the lowest decay rate occurs near $\delta=E_+ +E_F/2$.
Notably, as illustrated in Fig.~\ref{fig:fig5s}(d), the decay rate is density dependent, and in most cases, we have $\Gamma\ll E_F$, which suggests that the repulsive polaron is a well-defined quasi-particle throughout the OFR. We also note that further into the BCS side with large enough $\delta$, the repulsive polaron branch would eventually enter the molecule-hole continuum as illustrated in Fig.~\ref{fig:fig2s}, where the repulsive polaron would become unstable. However, at large $\delta$ but before the repulsive polaron branch merges into the molecule-hole continuum, the decay rate appears to decrease with increasing $\delta$, which is due to the decreasing quasi-particle residue $Z_+$ at large $\delta$.

\section{Phase separation}\label{sec_ps}
One of the reasons for the recent interest in repulsive polarons is the potential existence of itinerant ferromagnetism in repulsively interacting two-component fermions. Previous theoretical studies have shown that itinerant ferromagnetism may be stabilized for alkali fermionic atoms in the repulsive branch~\cite{reppol1,reppol2,reppol3,reppol4}. However, a direct experimental confirmation is still lacking. A natural question is whether itinerant ferromagnetism exists in alkaline-earth(-like) atoms under the spin-exchange interactions near an OFR.

\begin{figure}[tbp]
\centering
\includegraphics[width=7.5cm]{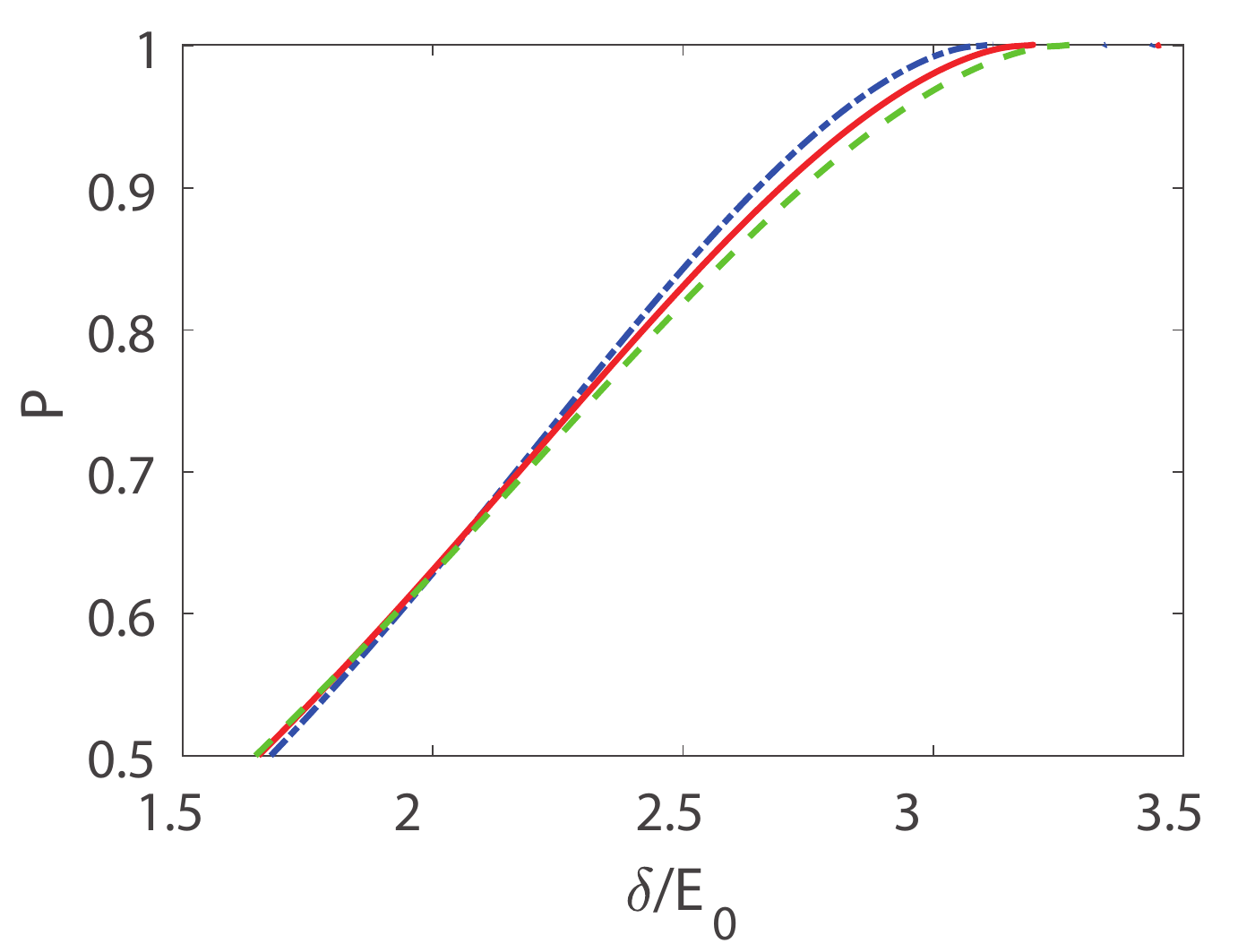}
\caption{Phase diagrams in terms of the detuning $\delta$ and the polarization $P=(N_1-N_2)/(N_1+N_2)$. Above and to the left of the line is the mixed phase, while below and to the right of the line is the phase-separated state. The green dashed, the red solid, and the blue dash-dotted curves correspond to $n=1.2n_0$, $n_0$, and $0.8 n_0$, respectively.}
\label{fig:fig6s}
\end{figure}

In an effort to answer this question, here we consider a system of $N_2$ impurity atoms of the state $|e,\uparrow\rangle$ immersed in $N_1$ majority atoms of the state $|g,\downarrow\rangle$. Following the treatment in Ref.~\cite{reppol4}, we derive the free energy of a homogeneous mixture and study the condition for the occurrence of a phase separation. For a highly polarized mixture with $y=N_2/(N_1+N_2)\ll1$, the energy per particle for a homogeneous mixture at zero temperature can be written as
\begin{equation}\label{Emix}
  E_{\rm mix}=\frac{3}{5}E_{F}^{(1)}(1-y)+\frac{3}{5}E_{F}^{(2)}y+yE_{+}(N_{1},\delta),
\end{equation}
where $E_{F}^{(1)}$ and $E_{F}^{(2)}$ are the Fermi energies of the atoms in the $|g,\downarrow\rangle$ and the $|e,\uparrow\rangle$ states, respectively. The last term $E_{+}(N_{1},\delta)$ in the expression above is the energy of a single $|e,\uparrow\rangle$ impurity atoms interacting with $N_1$ $|g,\downarrow\rangle$ atoms in the repulsive branch, which can be related to the repulsive polaron energy as $E_{+}(N_{1},\delta)=E_+(1-y)^{2/3}$, where $E_+$ is the repulsive polaron energy discussed in the previous sections with a total particle number $N=N_1+N_2$. By applying the usual Maxwell construction to $E_{\rm mix}$~\cite{reppol4}, we obtain the critical polarization from the minimum of $E_{\rm mix}$. In Fig.~\ref{fig:fig6s}, we plot the resulting phase diagram on the plane of detuning $\delta$ and polarization $P \equiv (N_1 - N_2)/(N_1 + N_2)$. Note that in the large polarization limit $P \sim 1$, the mixture becomes unstable towards phase separation beyond a critical $\delta_c$, which is density dependent due to the non-universal nature of the OFR.

\begin{figure}[tbp]
\centering
\includegraphics[width=7.5cm]{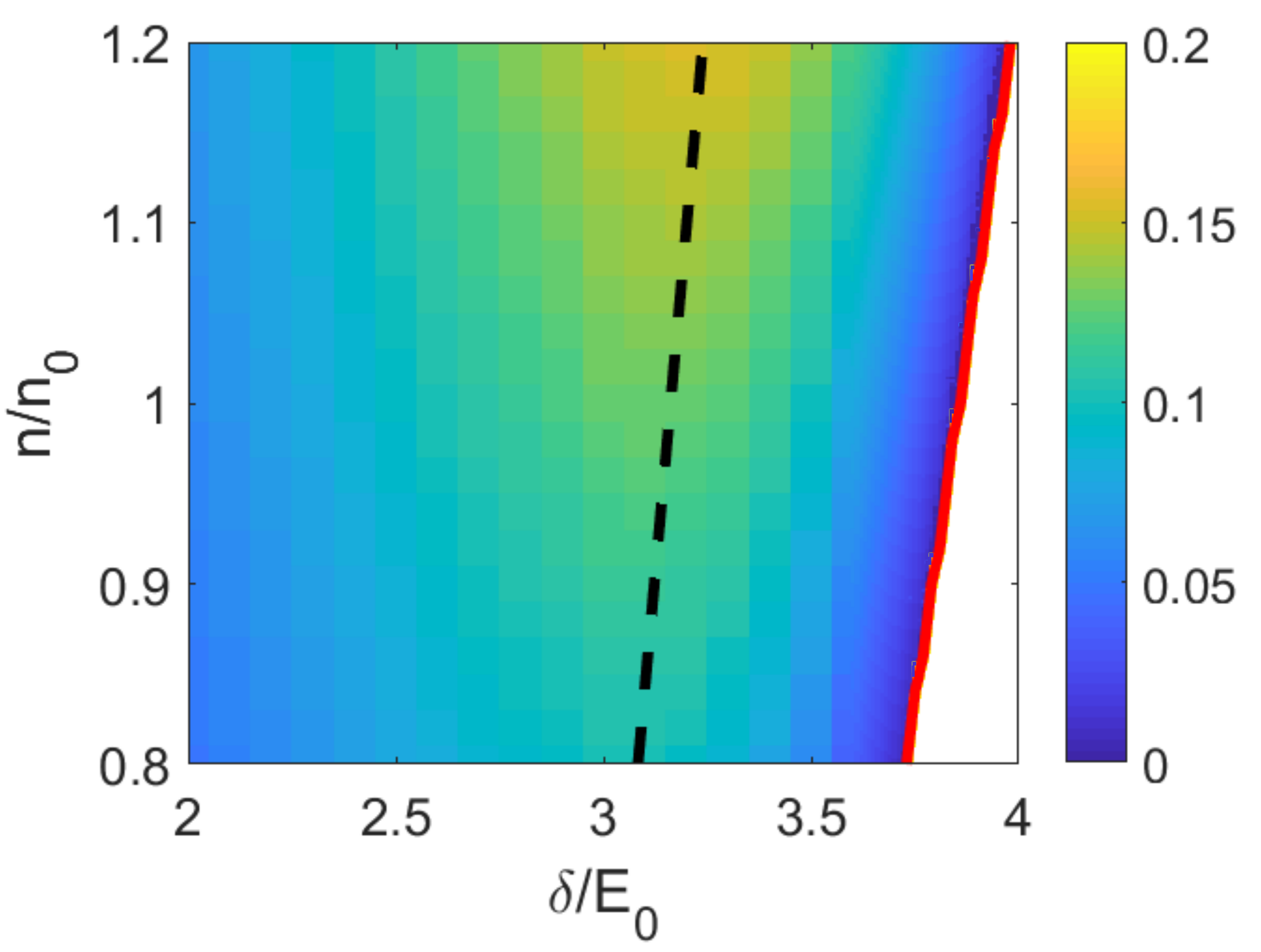}
\caption{Phase diagram for the stability of the phase-separated state on the $\delta$-$n$ plane. The background color shows the decay rate $\Gamma/E_0$ of the repulsive polaron. The black dashed line indicates the phase boundary between the mixed and the phase-separated states. The red solid line is the boundary where the repulsive polaron merges into the molecule-hole continuum. Here we take the large polarization limit with $P=1$.}
\label{fig:fig7s}
\end{figure}

The stability region of the phase-separated state shown in Fig.~\ref{fig:fig6s} is conditional on the stability of the repulsive polaron. More specifically, the repulsive polaron should be a long-lived, well-defined quasi-particle away from the molecule-hole continuum to justify the free-energy considerations leading to the phase diagram in Fig.~\ref{fig:fig6s}. To provide further perspective on this point, we show in Fig.~\ref{fig:fig7s} the phase diagram in the large polarization limit $P=1$ on the $\delta$--$n$ plane, against the false color plot of the polaron decay rate. We conclude that the phase-separated state, and hence the itinerant ferromagnetism, may be stabilized to the immediate left of the red-dotted line, where the decay rate $\Gamma<0.1E_0$ and the phase-separated state is energetically favorable.

\section{Summary}\label{sec_sum}
We have characterized in detail the key properties of the repulsive polaron near an OFR, using the parameters of $^{173}$Yb. We find that the two-channel nature of the OFR has significant impact on the properties of the repulsive polaron. In particular, the decay rate features a minimum at small magnetic field, on the BEC side of the resonance point. The dressing of the repulsive polaron by the closed-channel scattering states would also give rise to visible kinks in both the residue and the effective mass of the repulsive polaron. We also estimate the parameter region where the itinerant ferromagnetism may be stabilized and observed. Our results can be readily checked using the exiting experimental techniques in alkaline-earth(-like) atoms.

\section*{Acknowledgments}
We thank Pietro Massignan and Zhenhua Yu for helpful comments and discussions. This work is supported by the National Key R\&D Program (Grant No. 2016YFA0301700), the NKBRP (2013CB922000), the National Natural Science Foundation of China (Grant Nos. 60921091, 11274009, 11374283, 11434011, 11522436, 11522545, 11774425), and the Research Funds of Renmin University of China (10XNL016, 16XNLQ03). W.Y. acknowledges support from the ``Strategic Priority Research Program(B)'' of the Chinese Academy of Sciences, Grant No. XDB01030200.

\end{document}